\documentclass[twocolumn]{aastex62}
\usepackage{amsmath}
\usepackage{rotating}
\usepackage{graphicx}

\newcommand{\gerg}{Gerg s$^{-1}$ cm$^{-2}$}
\begin{document}
\title{Slow Cooling and Fast Reinflation for Hot Jupiters}
\author[0000-0002-5113-8558]{Daniel P. Thorngren}
\affil{Department of Physics, University of California, Santa Cruz}
\affil{Institute for Research on Exoplanets (iREx), Universit\'e de Montr\'eal, Canada}
\author[0000-0002-9843-4354]{Jonathan J. Fortney}
\affil{Department of Astronomy and Astrophysics, University of California, Santa Cruz}
\author[0000-0002-7727-4603]{Eric D. Lopez}
\affil{NASA Goddard Space Flight Center, 8800 Greenbelt Rd, Greenbelt, MD 20771, USA}
\author[0000-0002-2580-3614]{Travis A. Berger}
\affiliation{Institute for Astronomy, University of Hawai`i, 2680 Woodlawn Drive, Honolulu, HI 96822, USA}
\author[0000-0001-8832-4488]{Daniel Huber}
\affiliation{Institute for Astronomy, University of Hawai`i, 2680 Woodlawn Drive, Honolulu, HI 96822, USA}
\begin{abstract}
The unexpectedly large radii of hot Jupiters are a longstanding mystery whose solution will provide important insights into their interior physics.  Many potential solutions have been suggested, which make diverse predictions about the details of inflation.  In particular, although any valid model must allow for maintaining large planetary radii, only some allow for radii to increase with time.  This reinflation process would potentially occur when the incident flux on the planet is increased.  In this work, we examine the observed population of hot Jupiters to see if they grow as their parent stars brighten along the main sequence.  We consider the relation between radius and other observables, including mass, incident flux, age, and fractional age (age over main sequence lifetime), and show that main sequence brightening is often sufficient to produce detectable reinflation.  We further argue that these provide strong evidence for the relatively rapid reinflation of giant planets, and discuss the implications for proposed heating mechanisms.  In our population analysis we also find evidence for a ``delayed-cooling effect'', wherein planets cool and contract far more slowly than expected.  While not capable of explaining the observed radii alone, it may represent an important component of the effect. Finally, we identify a weak negative relationship between stellar metallicity and planet radius which is presumably the result of enhanced planetary bulk metallicity around metal-rich stars and has important implications for planet formation theory.
\end{abstract}

\section{Introduction}
Since the first discovery of a transiting hot Jupiter, HD 209458 b \citep{Charbonneau2000,Henry2000}, there has been an open question as to why their radii are so large.  Applying only the physics seen in Jupiter and Saturn, giant planets should not have radii much exceeding $\sim1.2 R_J$ \citep{Fortney2007}, even on close orbits.  Nevertheless it has  become clear that hot Jupiters almost always have inflated radii, extending to as much as twice the radius of Jupiter \citep[e.g.][]{Hartman2011c}.  Since then, a great deal of research has been conducted to identify the missing physics that could explain this discrepancy.

Many solutions have been suggested.  \citet{Bodenheimer2001} note that tidal dissipation, driven by interactions with other planets in the solar system, would deposit significant energy into the planetary interior.  \citet{Batygin2010, Perna2010a, Wu2013} suggest that Ohmic dissipation in the atmosphere could transfer heat downwards, heating the interior and suppressing heat loss.  \citet{Arras2009} propose that tidal forces acting on planets' thermal bulges could deposit heat into the interior.  \citet{Chabrier2007} and \citet{Burrows2007} theorize that large composition gradients or additional atmospheric opacity sources respectively could slow interior cooling to a crawl.  Fluid dynamical effects might also allow heat to be pushed from the atmosphere into the interior \citep[e.g.][]{Guillot2002,Youdin2010,Tremblin2017}.  Yet this is still only a sample of the many proposed solutions.

To sort through these theories, we must identify testable differences between them.  One valuable piece of evidence is that the radius excess appears to only occur at equilibrium temperatures exceeding 1000 K \citep{Miller2011,Demory2011}.  Similarly, these radius anomalies correlate with incident flux much better than semi-major axis \citep{Weiss2013,Laughlin2011}, disfavoring tidal explanations which would scale against the latter.  In \cite{Thorngren2018}, we showed that the extra interior heating as a fraction of flux (fit to match the observed radii) decreases at very high fluxes.  This was predicted by the Ohmic dissipation model \citep[see][]{Menou2012,Batygin2011}, but may be consistent other models as well.

Another test, proposed in \cite{Lopez2016}, would determine whether hot Jupiters reinflate when their insolation increases by identifying hot Jupiters orbiting red giants, whose insolation greatly increased when their stars evolved off the main sequence.  Whether they would reinflate depends on how deeply the anomalous heat is deposited.  \cite{komacek2017a} showed that heating below the radiative-convective boundary allows planets to remain inflated, but heating above this point has little effect.  However, in order to reinflate planets the heating must be deposited very deep \citep{komacek2020}, as the energy can only flow downward very slowly via conduction -- convection is fast, but only moves heat upward.  Tidal heating mechanisms, such as thermal tides \citep[e.g.]{Arras2009,Socrates2013,Gu2019} may deposit heat deep enough to achieve this, but neither delayed cooling models nor Ohmic dissipation \citep{Ginzburg2016} can do so on their own.  Some planets orbiting red giants have been discovered \citep{Grunblatt2016,Grunblatt2017} that seem to point towards reinflation occurring, but the results are not yet conclusive.

Of particular interest to us is a companion path to the \cite{Lopez2016} approach discussed in \cite{Hartman2016}, which found a correlation between the radius of hot Jupiters and the fractional age of their parent star (age divided by main sequence lifetime).  Their explanation was that as stars brighten over time on the main sequence, the flux on the hot Jupiters increases in kind and their radii grow.  They showed that their results were the same for HAT, WASP, and Kepler planets separately as well as for the combined population.  However, they did not attempt to control for other observables, such as mass, eccentricity, or stellar mass.  This is important to show that it is not simply a byproduct of some other correlation.  Additionally, \cite{Hartman2016} predates the release of Gaia DR2 data, so we have a chance to revisit the issue with this additional, high-quality data.

For this work, we will further investigate the possibility of main-sequence reinflation of giant planets.  First we will study which variables are predictive of the planet radius under a variety of control schemes (\S\ref{sec:massFluxRadius}).  Next, we will show that the main-sequence brightening is indeed strong enough to reinflate some planets detectably -- and thus that it is a good test of whether reinflation occurs (\S\ref{sec:reinflationEffects}).  At last, we will consider the evidence from hot Jupiters around main-sequence stars that this reinflation actually occurs (\S\ref{eq:massFluxRadius} and \ref{sec:radiusAge}).  Along the way, we will uncover evidence that stellar metallicity predicts smaller planet radii (\S\ref{sec:massFluxRadius}) and that a delayed cooling effect affects the radii of younger hot Jupiters (\S\ref{sec:radiusAge}).

\section{Data Collection} \label{sec:data}
Our exoplanet parameters were collected by combining data from exoplanet.eu \citep{Schneider2011} and the NASA Exoplanet Archive \citep{Akeson2013}.  We then combed through the results using both automated consistency checks and manual examination to join duplicate listings (e.g. WASP-183 A b and KELT-22 A b) and remove planets whose radii were determined theoretically (e.g. 51 Peg b).  We limit our investigation to hot Juptiers above the empirical inflation limit at $F > 2\times 10^8$ \gerg \citep{Miller2011,Demory2011} with observed masses $M > 0.5 M_J$ to avoid including hot Saturns, which seem to exhibit a different relationship between flux and radius \citep[see][]{Thorngren2018}.

In this study, we are particularly concerned with the stellar age and luminosity history, and so have expended additional effort to ensure this data was of a high quality.  We collected spectroscopic metallicity and effective temperatures from SWEET-Cat \citep{Santos2013}, NASA Exoplanet Archive \citep{Akeson2013}, and occasionally directly from the source papers \citep[e.g.][]{Bonomo2017, Livingston2018, Espinoza2019}, taking care to ensure that both values were derived from the same source.  These were combined with Gaia Data Release 2 parallaxes \citep{GaiaCollaboration2018} and 2MASS \cite{Skrutskie2006} K magnitudes and fed into the stellar fitting code IsoClassify \citep{Huber2017, Berger2020}.  In cases where the K magnitude was not available, we used the Gaia G magnitude instead, accounting for the color differences between these bandpasses.  This allows us to uniformly compute stellar ages which account for Gaia information (unlike most original discovery estimates) and self-consistent fractional ages.  It also provides us with stellar luminosity histories, which we use to determine the maximum possible inflation since the system formed.  In total, our data consists of 318 hot Jupiters and their host star parameters.

\section{Mass, Flux, and Radius}\label{sec:massFluxRadius}
To understand how the radius of a hot Jupiter is set, it is worth studying which observables correlate with radius in a comprehensive manner free of interior modelling assumptions.  To do this, we will compare the Bayesian Information Criterion (BIC) for various linear regressions of the log of the parameters.  Specifically we consider simple linear models of the following form:
\begin{equation}
    \log(R) \sim \mathcal{N}\left(
        \beta_0 + \sum_{i=1}^n \beta_i \log(x_i),
    \sigma \right)
    \label{linearModel}
\end{equation}

Where $\mathcal{N}$ is the normal distribution, $x_i$ are the observed regressors under consideration, $\beta_i$ are the regression coefficients, and $\sigma$ is the residual standard deviation.  We use the standard reference priors $p(\vec{\beta},\sigma) \propto \sigma^{-2}$.  This reduces to a power-law relationship for $R$, but writing it as a linear model lets us use the least-squares regression solution.  Because we have such a simple model, the BIC is also available in a closed form.  Let $n$ be the number of planets, $k$ the number of regressors ($x$), and SSD the sum of squared residuals.  Then up to a constant term, the BIC is:
\begin{equation}
    \mathrm{BIC} = k\log(n) + n(1 + \log(\mathrm{SSD}/n))
\end{equation}

We will consider the following regressors: the planet mass $M$ (in $M_J$), the incident flux on the planet $F$ (in \gerg), the stellar mass $M_\star$ (in $M_\odot$), the stellar radius $R_\star$ (in $R_\odot$), the system age $t$ (in Gyr), the fractional age $t/T$ (age over main sequence lifetime), the orbital semi-major axis $a$ (in AU), the orbital eccentricity $e$, stellar luminosity $L_\star$, the stellar zero-age-main-sequence (ZAMS) flux $F_\mathrm{zams}$ (in \gerg), and the second order crossterm $\exp(\log(M)\log(F))$.  The crossterm was included to allow for the effect seen in structure models where higher mass planets are more difficult to inflate for an equal amount of heating \citep[see][]{Thorngren2018}.  For $t$ and $t/T$, we omit the log in Eq. \ref{linearModel} to best reflect the potential relationship, and likewise for $e$ and [Fe/H] to avoid taking the log of zero.

\begin{sidewaystable}
    \centering
    \begin{tabular}{lrlllllllrllll}
\hline
 Base Model                                                       &   $\sigma/R$ & $M$            & $F$             & $M_\star$      & $R_\star$      & $t$            & $t/T$         & $a$           &   $e$ & $L_\star$      & $F_\mathrm{zams}$           & $\mbox{[Fe/H]}$   & $[MF$$]$       \\
\hline
 1.26                                                             &        18.66 & 2.8            & \textbf{-116.0} & \textbf{-65.6} & \textbf{-60.1} & \textbf{-28.6} & \textbf{-6.4} & 3.5           &  2.50 & \textbf{-74.4} & \textbf{-115.2}     & \textbf{-2.3}     & 2.4            \\
 1.21$F^{0.11}$                                                   &        15.20 & \textbf{-18.1} &                 & 2.3            & 3.7            & 2.8            & 5.8           & 3.0           &  3.20 & 0.9            & 3.2                 & \textbf{-0.7}     & \textbf{-15.0} \\
 1.22$M^{-0.06}$$F^{0.12}$                                        &        14.63 &                &                 & 3.0            & 4.7            & 2.8            & 5.2           & 3.1           &  5.20 & 1.9            & 3.0                 & \textbf{-3.2}     & \textbf{-3.5}  \\
 1.22$M^{-0.04}$$F^{0.14}$$[MF$$]^{-0.07}$                        &        14.43 &                &                 & 2.8            & 5.2            & 2.1            & 4.1           & 3.4           &  4.50 & 2.0            & \textbf{-0.2}       & \textbf{-4.6}     &                \\
 1.24$M^{-0.04}$$F^{0.14}$$[MF$$]^{-0.08}$$\mbox{[Fe/H]}^{-0.06}$ &        14.20 &                &                 & \textbf{-2.0}  & 4.3            & 1.2            & 3.6           & 3.0           &  5.00 & 1.6            & 1.2                 &                   &                \\
 1.28$F_\mathrm{zams}^{0.11}$                                     &        15.22 & \textbf{-16.9} & 2.5             & \textbf{-0.8}  & \textbf{-3.5}  & 5.5            & \textbf{-0.2} & \textbf{-3.8} &  5.50 & \textbf{-5.6}  &                     & 1.0               & \textbf{-20.4} \\
\hline
 1.22                                      &        17.85 & 4.1   & \textbf{-35.8} & \textbf{-26.1} & \textbf{-32.3} & \textbf{-7.7} & \textbf{-11.0} & 4.4           &  3.30 & \textbf{-31.7} & \textbf{-34.6}      &              1.50 & \textbf{-0.7} \\
 1.21$F^{0.11}$                            &        14.44 & 0.3   &                & 0.8            & \textbf{-1.8}  & 3.9           & 0.1            & \textbf{-1.5} &  4.60 & \textbf{-1.5}  & 4.0                 &              1.70 & 1.5           \\
 1.22$M^{-0.04}$$F^{0.12}$                 &        14.19 &       &                & 2.7            & 0.5            & 4.5           & 1.0            & 0.9           &  4.40 & 0.9            & 4.5                 &              1.50 & 2.3           \\
 1.23$M^{-0.04}$$F^{0.14}$$[MF$$]^{-0.06}$ &        14.09 &       &                & 2.0            & \textbf{-0.3}  & 4.4           & 1.0            & 0.1           &  4.30 & 0.1            & 4.6                 &              1.50 &               \\
\hline
 1.19                         &        16.90 &  2.90 & \textbf{-33.1} & \textbf{-6.8} & \textbf{-10.3} &  2.50 & \textbf{-6.7} &  1.80 &  3.40 & \textbf{-13.5} & \textbf{-30.6}      & \textbf{-5.5}     &       3.60 \\
 1.18$F^{0.09}$               &        15.04 &  0.80 &                & 5.1           & 4.9            &  5.00 & 4.6           &  4.90 &  4.40 & 4.4            & 4.9                 & \textbf{-2.9}     &       2.60 \\
 1.25$F_\mathrm{zams}^{0.09}$ &        15.16 &  1.40 & 2.4            & 3.9           & 1.3            &  4.10 & \textbf{-1.3} &  1.40 &  4.90 & 0.4            &                     & \textbf{-2.1}     &       1.50 \\
\hline
\end{tabular}
    \caption{A table of $\Delta$BICs showing which variables are effective in predicting the radius as we include progressively more variables in a power-law model.  Negative numbers (in bold) indicate that the BIC criterion favors adding the variable to the model.  $\sigma/R$ is the relative predictive uncertainty as a percent.  Columns are described in the text, but note especially that $[MF]$ is the next-order crossterm between $M$ and $F$, not the product of $M$ and $F$.  The table is divided into three blocks.  The top block was calculated for the full set of 318 planets, whereas the middle block was calculated only for planets for whom $F/F_0 < 1.5$: these 100 planets could have reinflated at most only a small amount (about 6\% in radius).  The bottom block shows the 172 planets older than 3 Gyr., to exclude planets that might still be cooling to equilibrium.}
    \label{tab:bics}
    
\end{sidewaystable}

We carried out this procedure for a number of models and regressors; Table \ref{tab:bics} shows the results.  Each row corresponds to a statistical model of the observed radius (on the left) with $\sigma/R$ as the residual uncertainty as a percent.  The remaining columns are the $\Delta BIC$ values for adding the regressor indicated by the column head.  If a variable is already in the model, its $\Delta BIC$ is omitted.  More negative numbers are indicate more favored additions to the model -- positive numbers indicate it is preferred to leave the variable out.  We have grouped the rows into three blocks: the full set of 318 planets in the top block, the subset of 100 for which the flux has increased by less than 50\% during the stellar lifetime, and the subset of 172 that are older than 3 Gyr.  The second group can act as a control group against reinflation, as the flux change is insufficient to bring about much reinflation under any of the models (see section \ref{sec:reinflationEffects}).  The final group was selected to remove planets that might still be cooling to equilibrium.  Of course, by reducing the size our data set, these cuts also reduce the significance of detected trends.

For example, in the second row we consider modifications to the base model $R=1.21F^{0.11}$.  Most of the variables considered would not improve this model if included (their $\Delta BIC$ is positive), but mass, [Fe/H] (marginally), and the mass-flux crossterm would.  The $M$ column of this row indicates that the a model of the form $R=c_0 F^{c_1} M^{c_2}$ (where $c_i$ are fitting constants) outperforms the row's base model with a $\Delta BIC$ of -18.1.  The next row considers this modified model as it's base, and we see that indeed the relative uncertainty $\sigma/R$ drops modestly from 15.20 to 14.63.  Such an improvement is not wildly important for accurate predictions, but the fact that it was statistically significant (as determined by the $\Delta BIC$) tells us that mass is somehow tied to the radius.  In this case, the physical cause of that connection is almost certainly that more massive planets have higher gravity to counteract the effect of a constant heating on the planet's radius \citep[see][Fig. 2]{Thorngren2018}.

It is clear from this analysis that flux is by far the most important variable.  When accounting for flux, it is followed distantly by the planet mass, and then the cross-term of $M$ and $F$.  After this, the stellar metallicity predicts slightly smaller planetary radii.  This might be caused by higher stellar [Fe/H] predicting higher planet metallicities, which would tend to shrink the planet radius.  We will test this correlation rigorously in \S\ref{sec:uncertainties}.

Furthermore, we see the correlation between fractional age and radius noted by \cite{Hartman2016}, and observe that the correlation vanishes when the present-day flux is accounted for.  This is consistent with their interpretation that hot Jupiters further reinflate as their parent stars age and brighten.  In support of this, when we only control for the flux predicted for the star at ZAMS, the fractional age is not fully eliminated.  This effect is more clear when comparing the two cases in stars older than 3 Gyr. are found (Table \ref{tab:bics}, bottom block).  This further supports the idea that it is serving as a proxy for the amount of main-sequence brightening (and therefore radius reinflation) that has occurred.

From these results we can also suggest a well-motivated model for the radius given other information.  We favor a model of the form
\begin{equation}\label{eq:massFluxRadius}
    % R = 1.22M^{-0.042} F^{0.137-0.072\log_{10}(M)},
    R = A M^{B} F^{C+D\log_{10}(M)}
\end{equation}
as this includes all of the significant variables -- we choose not to include [Fe/H] as the predictive improvement is marginal.  Our log-linear fits to the whole data set (Table \ref{tab:bics}) give values for the constants $A=1.22$, $B=-.042$, $C=.137$, and $D=-.072$.  If we consider only planets whose flux increased by $<50\%$ so far (the bottom block of Table \ref{tab:bics}), we can see that fitting for this model yields nearly identical results.  This supports the proposition that planetary radii react to the present-day flux rather than the full flux history.

\subsection{Uncertainty Analysis}\label{sec:uncertainties}
The statistical models discussed so far do not account for uncertainty in the observed planetary properties.  To ensure that this doesn't impact our results, we have run a hierarchical Bayesian regression on our key model in Eq. \ref{eq:massFluxRadius}, accounting for the uncertainties in mass and radius.  We sampled the posterior using a Gibbs sampler (see e.g. \citet{Gelman2014}, pg. 276) and verified convergence using the Gelman-Rubin diagnostic \citep{Gelman1992}.  We find a coefficient of $A = 1.21 \pm 0.01$ and exponents of $B=-0.045 \pm 0.01$, $C=0.149 \pm 0.0093$, and $D=-0.072 \pm 0.0213$, yielding a relative predictive uncertainty of $11.3\%$.  These results are very similar to those of our original fit for Eq. \ref{eq:massFluxRadius}, so we conclude that neglecting the observational uncertainties did not hamper that analysis significantly.

\begin{figure}[t]
    \centering
    \includegraphics[width=\columnwidth]{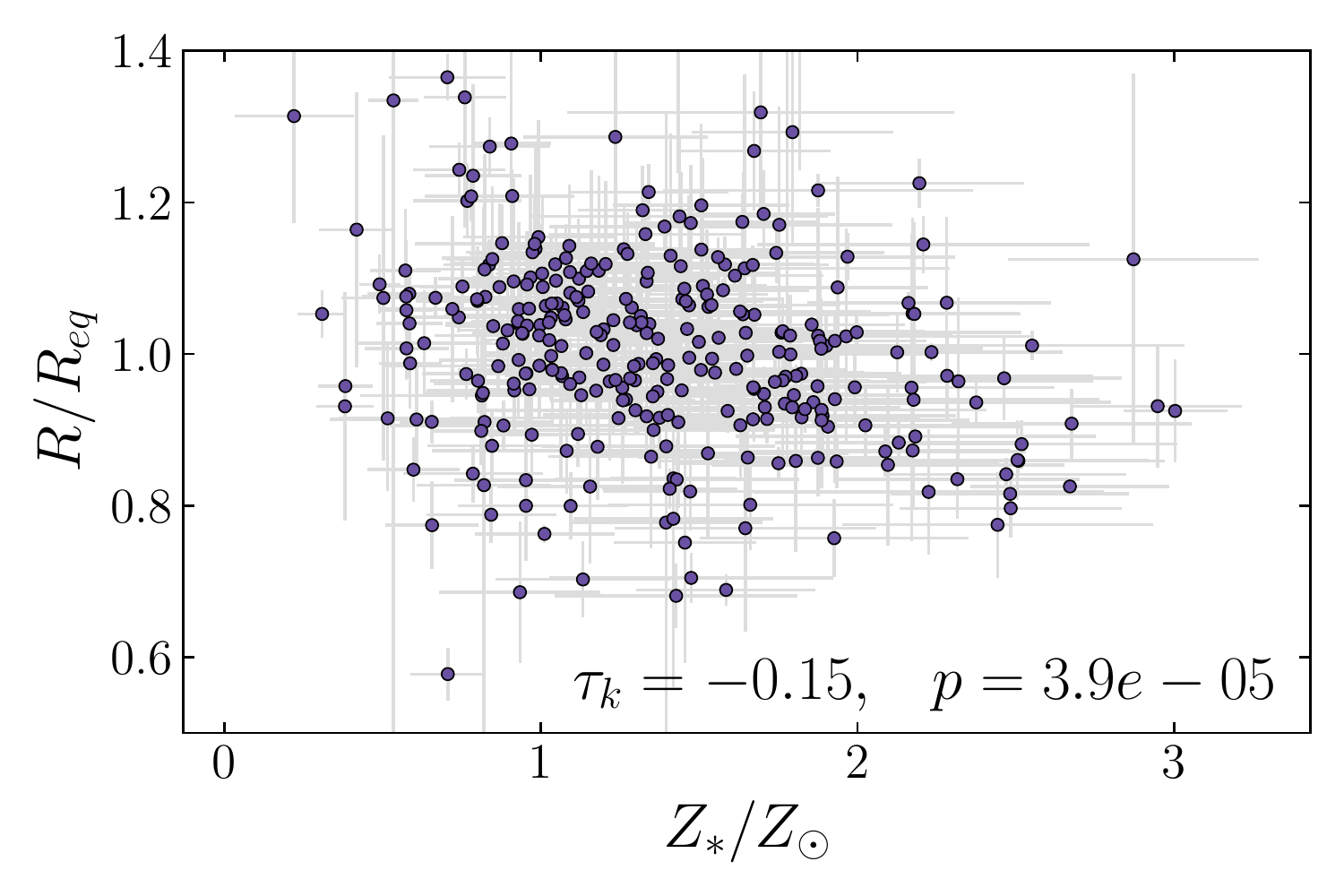}
    \caption{The observed radius relative to the expected radius from Eq. \ref{eq:massFluxRadius} (which we will call the equilibrium radius; see section \ref{sec:reinflationEffects}) plotted against the metallicity of the parent star.  A slight downward trend with significant scatter is apparent; such a trend would indicate that increased parent star metallicity correlates with higher planetary metallicity and thus smaller radii.  The $\tau_k$ and p-value shown are the results of a Kendall's Tau correlation test, which showed that the trend was strongly statistically significant.  In Section \ref{sec:uncertainties}, we discuss several additional statistical tests and a hierarchical Bayesian model which all find the effect to be statistically significant.}
    \label{fig:feResid}
\end{figure}

Our result that [Fe/H] corresponds to smaller radii also merits close examination as the relationship (Fig. \ref{fig:feResid}) exhibits significant scatter.  Our 318 planets had an average host star metallicity of 0.105 dex with standard deviation 0.177.  The median uncertainty was 0.097 dex.  We begin by testing for correlation with Kendall's tau ($\tau=-0.155$, $p=3.85\times10^{-5}$; see explanation in section \ref{fig:ageRadius}), Spearman's Rho ($\rho=-0.224$, $p=5.31\times10^{-5}$), and Pearson's r ($r=-0.180$, $p=0.00128$); all indicate a statistically significant, negative correlation.  To ensure the relation is not driven by outlier points, we also try a bootstrapped Kendall's Tau test, and find that just $3.9\times10^{-5}$ of the bootstrap samples do not exhibit a negative correlation ($\tau < 0$).  We get comparable results on all of these tests when we restrict our data to the 149 planets with low uncertainties ($\sigma_{[\mathrm{Fe/H}]} < 0.1$, $\sigma_{R/Req} < 0.1$).

To test the correlation while properly accounting for observational uncertainties, we conduct a hierarchical Bayesian fit similar to the one at the start of this section. This time we account for uncertainties in the mass, radius, and parent star metallicity.  The model we consider is the same as from Table \ref{tab:bics}, which to be explicit is:
\begin{align} \label{eq:feMassFluxRadius}
    \log_{10}(R) = &A + B \log_{10}(M) + C \log_{10}(F) + \nonumber\\
    & D \log_{10}(M)\log_{10}(F) + E [\mathrm{Fe/H}]
\end{align}
This is the linear form that allows for easy fitting; taking the base-10 exponential of both sides yields a power law in R.  Fitting the model exactly as in the previous paragraph, we find posterior distributions of $A =0.092 \pm 0.004$ ($1.24 \pm 0.011$ as a power-law form coefficient), $B = -0.047 \pm 0.011$, $C = 0.1445 \pm 0.0093$, $D = -0.0779 \pm 0.0212$, and (for the [Fe/H] term) $E = -0.0621 \pm 0.0175$, with a predictive uncertainty of $0.045$ (relative uncertainty of $10.9\%$ in power-law form).  Thus the [Fe/H] term works out to be non-zero to 3.2 sigma, even accounting for observational uncertainties, and we conclude that the result is statistically significant.  This has important implications for planetary composition and formation (see \S \ref{sec:conclusions}).

\section{The Effects of Reinflation}\label{sec:reinflationEffects}
\begin{figure*}[t]
    \centering
    \includegraphics[width=\textwidth]{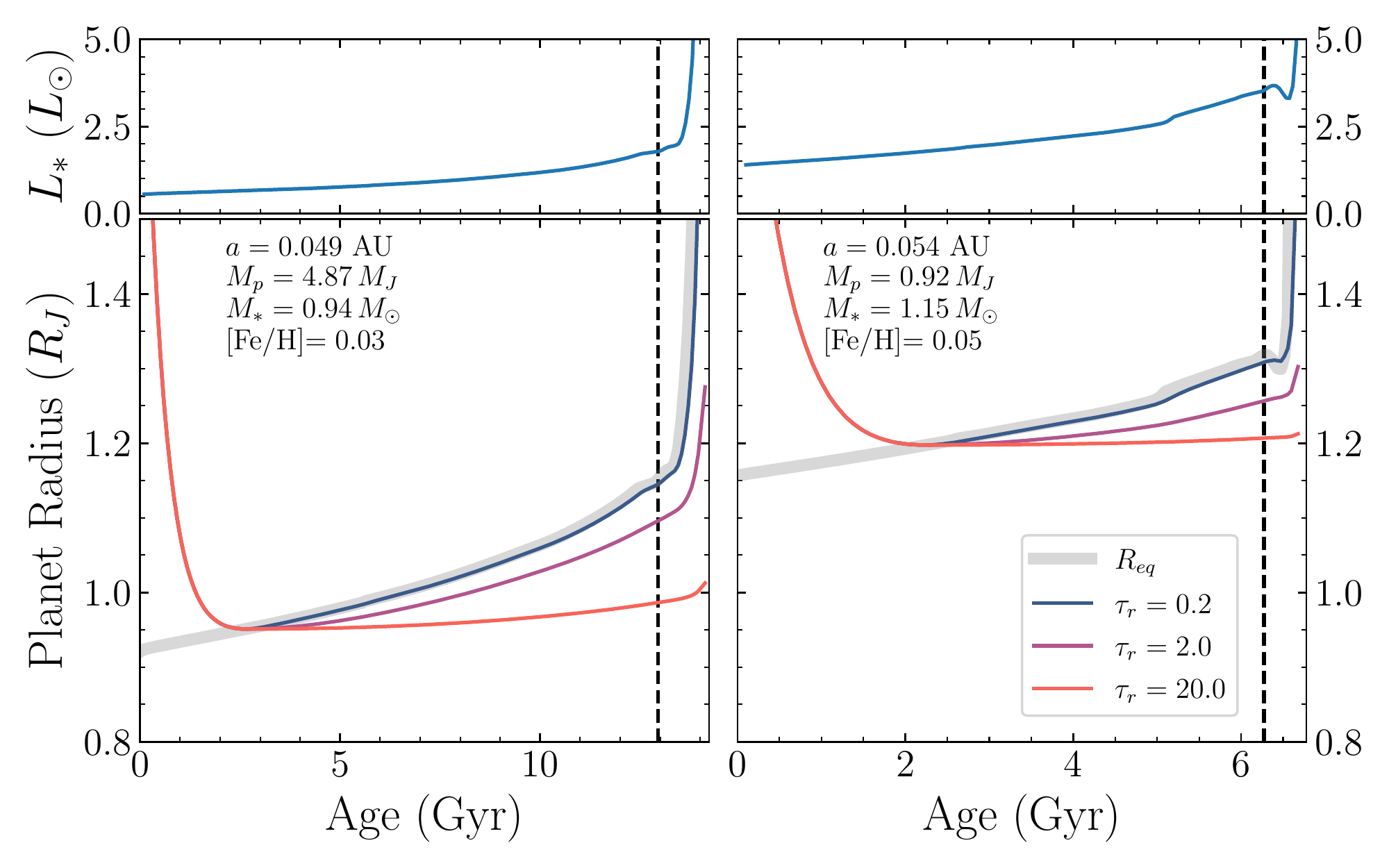}
    \caption{The luminosity evolution of stars (top row) and the resulting radius evolution of their planets (bottom row) for two different systems (columns).  The left column is based on HAT-P-21 b, and the right is based on HATS-55 b.  The thick grey line represents the equilibrium radius from Eq. \ref{eq:req}, the colored lines represent different values of the reinflation timescale $\tau_r$ in Gyr; we assume a deflation timescale of 0.5 Gyr.  The vertical dotted line is approximately the end of the main sequence; post main-sequence reinflation is apparent as well.  Stellar evolution tracks are from \cite{Dotter2016,Choi2016}.  Note that the equilibrium radius increases by more than 0.1 $R_J$ in both cases.}
    \label{fig:example}
\end{figure*}

Having surveyed the predictors of hot Jupiter radii, we will now verify that main-sequence stellar brightening is sufficient to inflate them to a detectable degree.  We will use Eq. \ref{eq:massFluxRadius} to predict the equilibrium radius $R_{eq}$ to which hot Jupiters would reach given enough time.  We could also choose to include stellar metallicity as in Eq. \ref{eq:feMassFluxRadius}, but trying this revealed no significant differences in our results, so we present the simpler model.  In principle, we could set these parameters by the slowly-evolving stars, but we saw in Table \ref{tab:bics} that this yields approximately the same fit regardless.  This equilibrium radius will vary as the flux changes with stellar evolution.  We will assume that real planets exponentially decay towards the equilibrium radius on some reinflation timescale $\tau_r$ when they are smaller than $R_{eq}$, and a separate deflationary timescale $\tau_d$ when they are larger than it.  Using the median parameter values from the Bayesian fit of \S\ref{sec:uncertainties}, this model takes the following form:

\begin{align}
    R_{eq}(t) = 1.21 M^{-0.045} F(t)^{0.149-0.072\log_{10}(M)} \label{eq:req}\\
    \frac{dR}{dt} = \frac{R_{eq}(t) - R}{\tau}\\
    \tau = \begin{cases}
        \tau_d & R > R_{eq}(t)\\
        \tau_r & R < R_{eq}(t)
    \end{cases} \label{eq:timescales}
\end{align}

There are a few cases to consider here.  First and simplest, if the timescale of reinflation $\tau_r$ is fast, then $R \approx R_{eq}$.  If the timescale of reinflation is very slow, then once $R < R_{eq}$, the radius becomes essentially constant.  For intermediate values of $\tau_r$, the situation is more complex.  These will increase slowly in radius as they approach a dynamical equilibrium where $dR/dt = dR_{eq}/dt$, assuming a steadily increasing stellar brightness.  However that condition will not always exist long enough for the planet to actually reach dynamical equilibrium.  This depends greatly on the rate of stellar brightening, the stellar lifetime, and the reinflation timescale.

\begin{figure*}[t]
    \centering
    \includegraphics[width=\textwidth]{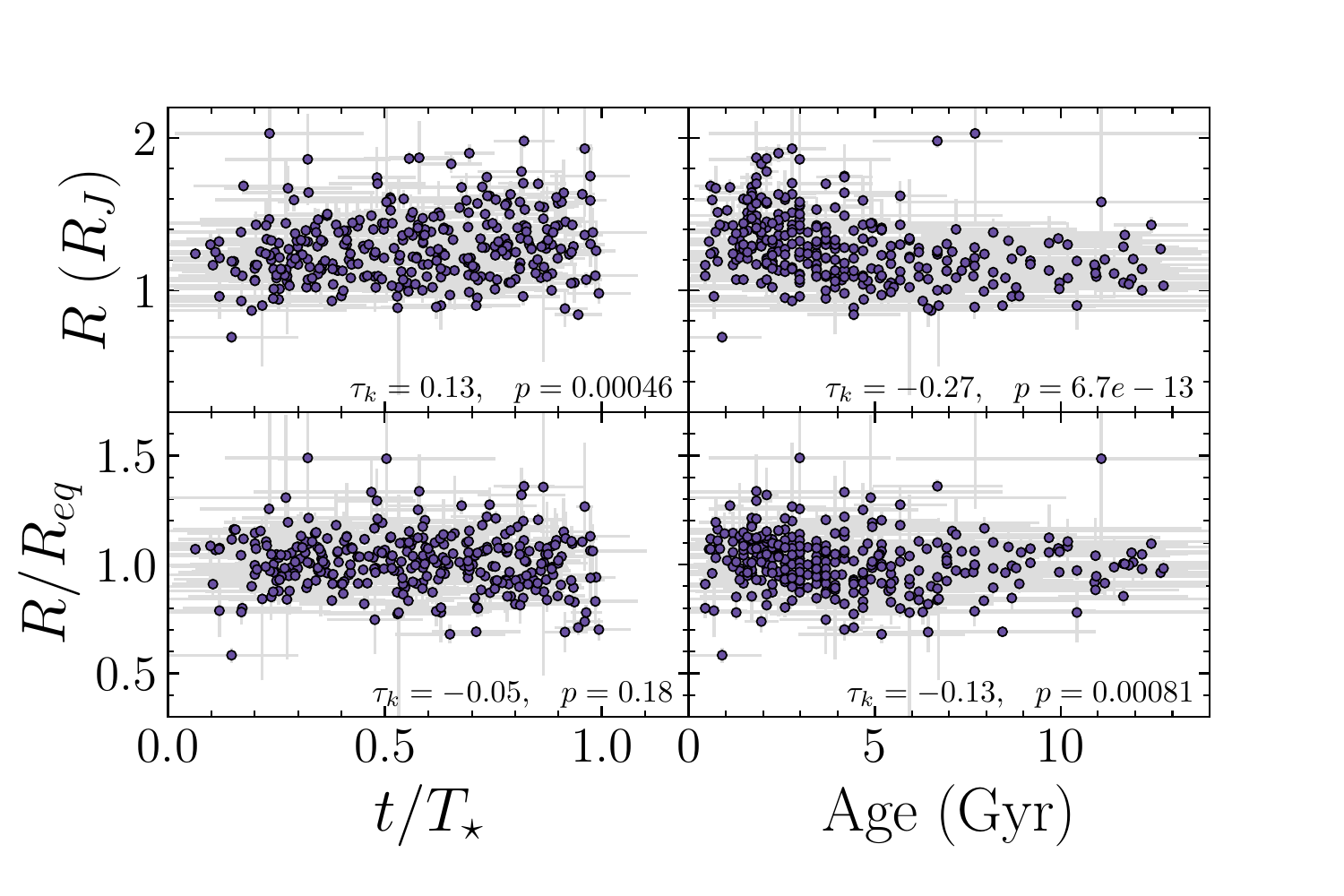}
    \caption{Planet radius (top) and the radius relative to the equilibrium radius (bottom, from Eq. \ref{eq:req}) against fractional age (left) and age (right).  To test for correlations, Kendall's tau ($\tau_k$) is shown along with the p-value for non-correlation null hypothesis.  A correlation exists between fractional age and radius (upper left), but not when flux and mass are corrected for (bottom left), consistent with fast reinflation.  A negative correlation with age exists in both cases, which appears in the plot as a downward trend in radius for the first few Gyr, followed by a leveling off.}
    \label{fig:ageRadius}
\end{figure*}

Using MIST stellar models \citep{Dotter2016,Choi2016} used in IsoClassify (see \S\ref{sec:data}), we can quantify the considerable brightening that stars undergo during their main-sequence lifetime.  Their relative luminosity changes are $\sim2\times$ for $1.5 M_\odot$ stars and $\sim4\times$ for $0.5 M_\odot$ stars (assuming solar metallicity).  Of course, the more massive stars experience this brightening much faster, so their relative brightening per unit time is greater.  The maximum possible reinflation (when $\tau_r \approx 0$) may be calculated from Eq. \ref{eq:req}, and for $M=1 M_J$ works out to approximately $5-20 \% $, depending on the stellar mass.  This should be readily detectable if $\tau_r$ is indeed short.

Using these luminosity tracks and solving equation equations \ref{eq:req} - \ref{eq:timescales}, we show radius evolution tracks in Figure \ref{fig:example} for several values of $\tau_r$ and two sets of system parameters.  In the left panel, based on HAT-P-12 b, the star brightens from $0.55 L_\odot$ at zero-age-main-sequence to $1.78 L_\odot$ at the end of its approximately 13 Gyr lifetime.  This results in $R_{eq}$ increasing from $0.92 R_J$ to $1.14 R_J$.  The colored lines show how the radius might keep up with this increase depending on the reinflation timescale; for a 20 Gyr timescale almost no reinflation occurs.  A similar effect is seen for a higher stellar mass case (based on HATS-55 b), but on a tighter schedule.  The stellar luminosity increases from $1.40 L_\oplus$ to $3.54 L_\oplus$ over about 6.25 Gyr, resulting in $R_{eq}$ increasing from $1.15 R_J$ to $1.31 R_J$.  Reinflation again could be detectable for fast timescales, but there is a smaller window towards the end of the star's lifetime when it would be detectable compared to the lower-mass case.

Both cases exhibit clear and detectable reinflation during the life time of the star when the reinflation timescale is short.  However, these predictions come with predictive uncertainty due to e.g. variations in planet composition and observational uncertainties.  Additionally, stellar luminosity tracks can be quite sensitive to the measured stellar properties.  Thus, we have found it to be unfeasible to directly fit a model of inflation and deflation to the observed data.  Instead we must rely on a careful examination of the population trends.  We have already explored some of this in Table \ref{tab:bics}, but it will be useful as well to reexamine the relationship between planetary radii and system age in detail.

\section{Radius and Age}\label{sec:radiusAge}
It is clear that reinflation is at least possible given the degree to which stars brighten on the main sequence.  Now we will examine whether the age-radius relationship is consistent with reinflation actually occurring.  Fig. \ref{fig:ageRadius} shows the radius and residual radius ($R/R_{eq}$ from Eq. \ref{eq:req}) plotted against the age and fractional age, with correlation measured by the Kendall's Tau statistic and its associated p-value.  This is a more flexible test of correlation than our approach in section \ref{sec:massFluxRadius}, as it can detect non-linear trends.  The statistic $\tau_k$ characterizes how well-sorted the data are in one variable after being sorted by the other variable; it runs from -1 (perfectly reverse sorted) to 1 (perfectly sorted).  The p-value is the standard statistical metric, with a conventional cutoff for significance at 0.05.  With high confidence ($p=4.6\times10^{-4}$) we are able to reproduce the positive correlation between radius and fractional age (top left panel) that \cite{Hartman2016} observed.  However, when we correct for the mass and incident flux using the equilibrium radius, the correlation vanishes ($p=0.18$, bottom left panel), as we saw in Table \ref{tab:bics}.  This supports fast reinflation; if its timescale were slow, the radius would instead lag behind $R_{eq}$ as the star evolved.

Interestingly, we also detect a \emph{negative} correlation in radius with the raw system age ($p=6.7\times10^{-13}$, top right panel).  Although this could be a sign of cooling to equilibrium among young planets, it could also relate to the stellar properties.  More massive stars are both brighter and live for less time, so one might find the most strongly insolated planets around them.  To separate this out, we examine the radius relative to $R_{eq}$ vs. the system age (bottom right panel).  Although the correlation does indeed weaken ($\tau_k = -0.13$ from $\tau_k=-0.27$), the negative correlation remains significant $p=8.1 \times10^{-4}$.  The bottom right panel of Fig. \ref{fig:ageRadius} shows radii of young hot Jupiters relative to $R_{eq}$ decrease with time, then level off.  This is evidence that hot Jupiters less than a few gigayears old are often still in the process of cooling.  Due to the massive internal temperatures required to explain their large radii, one would expect that these planets would cool rapidly; indeed, the \cite{Thorngren2018} models show that they should cool to equilibrium within the first 100 Myr at most.  That we see cooling occurring out to more than a Gyr implies that something is significantly inhibiting the cooling process (a delayed cooling effect).  However, this effect is still too short lived to explain the radii of older hot Jupiters, nor why the equilibrium radii are so large in the first place.  For that, additional internal heating is still required, as is the case generally for planet reinflation \citep{komacek2020}.

\section{Conclusions}\label{sec:conclusions}
Our results give strong support to the hypothesis that hot Jupiters can reinflate with their parent stars' main sequence evolution.  We have seen that the brightening of these stars, and the reaction of hot Jupiters to changes in flux, is significant enough that the resulting reinflation ($5-20\%$) would likely be detectable.  Whether this actually occurs depends on the reinflation timescale $\tau_r$ (see Fig. \ref{fig:example}); short timescales allow a planet to quickly reinflate along with its parent star's brightening, whereas long timescales would keep it largely static after cooling.  We find several of points of evidence that the reinflation timescale is fast:
\begin{enumerate}
    \item Planet radius is correlated with the fractional age of the parent star (reproducing the findings of \cite{Hartman2016}).
    \item The radius to fractional age correlation is completely eliminated by correcting for present-day flux, but \emph{not} when using the zero-age-main-sequence flux (see the last two rows of Table \ref{tab:bics}), suggesting that the fractional age is merely a proxy for stellar brightening on the main sequence.
    \item We find roughly the same solution for radius based on mass and flux (Eq. \ref{eq:massFluxRadius}) whether we fit to the full set of hot Jupiters or just those whose parent stars have not brightened enough to cause significant reinflation.  This means the present-day flux predicts a planet's radius without regard to the flux history because the radius is keeping up with flux changes.
    \item If reinflation were slow, planetary radii should lag behind equilibrium radii as the parent star evolves, but we do not observe this (Fig. \ref{fig:ageRadius}).
\end{enumerate}

In addition to our results on reinflation, we have also detected a pattern of radii larger than equilibrium among hot Jupiters in their first few gigayears (\S \ref{sec:radiusAge}).  This likely indicates that these planets are still in the process of cooling long after standard thermal evolution models would predict \citep[e.g.][]{Thorngren2018}.  This delayed cooling effect has been predicted elsewhere, perhaps as a result of internal composition gradients \citep[e.g.][]{Chabrier2007} or additional atmospheric opacity \citep{Burrows2007}.  However, it cannot be the sole driver of hot Jupiter inflation, as it predicts no reinflation.  Further, the cooling timescales we observe (Fig. \ref{fig:ageRadius}, bottom right panel) are too short to explain the radii of older planets.  However, delayed cooling may reduce the power required to maintain planets at their equilibrium radius, and so likely represent one part of the solution.

Fast reinflation also does not match predictions for the Ohmic dissipation model \citep{Batygin2010,Batygin2011,Menou2012}, which expect reinflation on a slow, 20 Gyr timescale \citep{Ginzburg2016}.  This is because its heating is deposited too shallow to quickly change the internal adiabat.  Indeed, any model which deposits its anomalous heating above the radiative-convective boundary will not produce much inflation \citep[see][]{komacek2017a} at all, and models must deposit their heat very deep within the planet to produce reinflation \citep{komacek2020}.  As such, Ohmic dissipation may be seen as broadly similar to delayed cooling models in its effect on planetary radii, and may be contributing to that effect.  It is not clear what fluid dynamical solutions like \citep{Guillot2002, Tremblin2017,Youdin2010} predict for reinflation; determining this will be a valuable test for these and future proposed explanations.  A combination of several mechanisms, such as that proposed in \cite{Sarkis2020}, appears to be a reasonable way to match our results.

Another interesting result of our analysis was that higher stellar [Fe/H] predicts smaller planetary radii, presumably because the planets have higher bulk metallicities.  The enhanced density of the resulting equation of state appears to affect the radius more strongly than any inhibition of cooling from enhanced atmospheric metallicity (and therefore opacity).  This result is also important because previous detections of this type of relationship were limited by the small available sample size \citep[e.g.][]{Guillot2006, Burrows2007}, or appear to have been the result of observational error \citep{Dodson-Robinson2012, Sarkis2015} and selection biases \citep{Dong2014,Gaidos2013a}.  Our analysis has the advantage of accounting for these uncertainties (see \S\ref{sec:uncertainties}) as well as correcting for the key predictors of mass and flux before considering the effect of stellar metallicity.  As such, this is the first statistically significant detection of a connection between planet radius and stellar metallicity when observational error is considered.  This result is particularly interesting in light of \cite{Teske2019}, which found that the relation between planet and star metallicity was weaker than a linear relationship.  Our results are consistent with this (the effect is small), but show that the relation is nevertheless nonzero.

There remains work to be done in this area.  In principle, it should be possible to directly fit thermal evolution models to the observed population.  To do this accurately, however, will necessitate carefully tying uncertainties in the stellar properties to the luminosity evolution, as well as considering how to represent both delayed cooling and internal heating in a generalized manner.  Second, the planets discovered with TESS are expected to include a large number of hot Jupiters \citep{Barclay2018}; these will provide a uniformly-derived sample on which to verify and expand upon our results.  Finally, the study of hot Jupiter hosting red giants proposed in \cite{Lopez2016} remains very much worthwhile as it could not only independently confirm our results but could also provide uniquely precise measurements of $\tau_r$.  This would require the discovery of more hot Jupiters around red giants \citep[as in][]{Grunblatt2017}, which would be interesting discoveries in their own right.

\acknowledgments
D. Thorngren acknowledges support by the Trottier Fellowship from the Exoplanet Research Institute (iREx).  JJF acknowledges the support of NASA Exoplanets Research Program grant NNX16AB49G.  T.A.B. and D.H. acknowledge support by a NASA FINESST award (80NSSC19K1424) and the National Science Foundation (AST-1717000). D.H. also acknowledges support from the Alfred P. Sloan Foundation.

\bibliography{bibliography}
\end{document}